\documentclass[epj]{svjour}
\usepackage{graphics}
\usepackage{epstopdf}
\usepackage{amsmath}
\usepackage{ulem}
\begin{document}
\title{Drag coefficient of a liquid domain in a two-dimensional membrane}
\author{
Sanoop Ramachandran\inst{1} 
\and 
Shigeyuki Komura\inst{1}
\thanks{E-mail: komura@tmu.ac.jp}
\and 
Masayuki Imai\inst{2}
\and 
Kazuhiko Seki\inst{3}
}                     

\institute{Department of Chemistry, 
Graduate School of Science and Engineering, 
Tokyo Metropolitan University, 
Tokyo 192-0397, Japan 
\and
Department of Physics, 
Faculty of Science, 
Ochanomizu University, 
Tokyo 112-0012, Japan
\and
National Institute of Advanced Industrial Science and Technology, 
Ibaraki 305-8565, Japan }
\date{Received: date / Revised version: date}
%
\abstract{
Using a hydrodynamic theory that incorporates 
a momentum decay mechanism, we calculate the drag coefficient 
of a circular liquid domain of finite viscosity moving in a 
two-dimensional membrane.
We derive an analytical expression for the drag coefficient which covers 
the whole range of domain sizes.
Several limiting expressions are  discussed. 
The obtained drag coefficient decreases as the domain viscosity 
becomes smaller with respect to the outer membrane viscosity.
This is because the flow induced in the domain acts to transport the 
fluid in the surrounding matrix more efficiently. 
} 

\maketitle

\section{Introduction}
\label{intro}

Membrane components such as lipids and proteins are subject to 
Brownian motion, and the resulting diffusive process is an 
important mechanism for the transport of these materials. 
Consequently, it is generally believed that many biochemical 
functions in membranes are diffusion controlled processes. 
Up to now, the studies of diffusion in biomembranes have mainly 
concentrated either on protein 
molecules~\cite{peters-82,lippincott-01,vrljic-02,reitz-01,gambin-06}
or, more recently, on lipid domains sometimes called as 
``lipid rafts''~\cite{klingler-93,oradd-05,cicuta-07,yanagisawa-07,sakuma-09}.
Lipid domains appear due to a lateral phase separation between a 
liquid-ordered phase and a liquid-disordered phase~\cite{veatch-05}, 
and have attracted much attention in relation to biological 
functionalities~\cite{simons-97}.

In the theoretical studies, a protein molecule is assumed to be a 
rigid disk moving in a two-dimensional (2D) thin fluid sheet under 
low Reynolds number conditions.
Saffman and Delbr\"uck (SD) investigated this problem by considering 
a membrane that is sandwiched in between a three-dimensional (3D) 
fluid medium \cite{saffman-75,saffman-76}.
They showed that the drag coefficient has only a weak logarithmic 
dependence on the disk radius in the small size limit.
The SD model was further analyzed by Hughes {\it et al.} for the whole 
range of protein sizes, and they also obtained the asymptotic expression 
in the large size limit~\cite{hughes-81}.
The SD model was extended by Levine \textit{et al.} for a viscoelastic film
within an infinite viscous liquid, taking into account both 
in-plane and out-of-plane dynamics~\cite{levine-02,levine-04,levine-04b}.

On the other hand, Evans and Sackmann looked at a slightly different 
situation~\cite{evans-88}, {\it i.e.}, the diffusion of a protein molecule 
moving in a supported membrane (instead of a free membrane).  
The presence of the solid substrate is taken into account through a 
friction term in the Stokes equation.
The equivalent hydrodynamic model was independently proposed by 
Izuyama who suggested that the momentum decay mechanism should 
generally exist for membranes surrounded by water~\cite{izuyama-88}. 
Since a hydrodynamic equation with a momentum decay term was  
originally proposed by Brinkman for flow in porous 
media~\cite{brinkman-47}, we call these approaches as the 2D 
Brinkman-type model.  
One of the advantages of the Brinkman model is that the drag coefficient 
can be analytically obtained over the whole range of disk sizes. 
In the small size limit, the SD model and the Brinkman model 
give the same logarithmic dependence.
In biological systems, the Brinkman model can be more relevant because 
the cell membranes are strongly anchored to the underlying cytoskeleton,
or are tightly adhered to other cells~\cite{alberts}.

Most of the previous work has concentrated on the problem of protein 
diffusion in a membrane for which the assumption of a rigid disk is 
reasonable.
Liquid lipid domains, on the other hand, cannot be considered as rigid
objects.  
A proper description should take into account the fluid nature and 
finite viscosity of the diffusing domain.
It should be also noted that the viscosity of the lipid domain is
different from that of the matrix~\cite{oradd-05}.  
In this paper, we derive the drag coefficient of a circular 
liquid domain by taking into account its finite viscosity using the 
2D Brinkman-type hydrodynamic equation. 
We show that the drag coefficient decreases as the domain viscosity 
becomes smaller.  
An analogous problem for a spherical drop of fluid was solved by 
Rybczy\'nski and Hadamard about 100 years 
ago~\cite{rybczynski-11,hadamard-11}.
Some authors considered the hydrodynamics of Langmuir monolayer domains 
at the air/water interface by taking into account the fluid nature 
of the domain~\cite{stone-95,lubensky-96,alexander-06}.  
In these works, however, the domain and the matrix viscosities are 
assumed to be the same, and the drag coefficient was not 
obtained~\cite{dekoker}.
Although the rotational diffusion coefficient was also calculated 
for a rigid disk~\cite{saffman-75,saffman-76,hughes-81,evans-88}, 
such a motion is irrelevant for a fluid domain which has an internal 
flow.

In the next section, we begin with a description of the 2D Brinkman model. 
In sect.~\ref{drag}, the solutions for the inside domain are connected 
to those for the outer membrane matrix through the appropriate boundary 
conditions.
Then the drag coefficient is calculated for the entire region of the 
domain sizes.    
We also obtain several limiting expressions; some of which coincide with 
the previous results.
Several discussions are provided in sect.~\ref{conclusion}.

\section{Hydrodynamic model with momentum decay}
\label{hydrodynamic}

The usual method of obtaining the diffusion coefficient $D$ from the 
drag coefficient $\zeta$ is to use the Einstein relation 
\begin{equation}
D = \frac{k_{\rm B}T}{\zeta},
\end{equation}
where $k_{\rm B}$ is the Boltzmann constant and $T$ the temperature.
However, for a pure 2D system in the low Reynolds number regime, a 
linear relation between the velocity and drag force cannot be 
obtained~\cite{landaulifshitz-flumech}.
This fact is due to the inability to simultaneously satisfy 
the boundary conditions both at the disk (cylinder) surface and at 
infinity within the Stokes approximation~\cite{lamb-75}. 
Such a problem is known as the Stokes paradox which essentially 
originates from the constraint of momentum conservation in a pure 
2D fluid.

Although a lipid bilayer membrane can be treated as a 2D viscous 
fluid, it is not an isolated system because the membrane is 
surrounded by a 3D fluid. 
Due to the coupling between the 2D membrane and the 3D fluid, the 
momentum within the membrane can leak away to the outer fluid.
Such an effect can be taken into account through a momentum 
decay term in the equation of motion.
This also enables the 3D fluid to be discarded and we can work in 
a pure 2D system.
A hydrodynamic equation which is consistent with the total momentum 
decay is 
\begin{equation}
\rho \left[ \frac{\partial \mathbf{v}}{\partial t} + 
(\mathbf{v} \cdot \nabla)\mathbf{v} \right]
= \eta \nabla^2\mathbf{v} - \nabla p - \lambda \mathbf{v}.
\label{hmeqn:6}
\end{equation}
In the above, $\mathbf{v}$ and $p$ are the 2D velocity and pressure, 
respectively, while $\rho$ is the 2D membrane density, $\eta$ the 
membrane 2D viscosity, and $\lambda$ the momentum decay parameter.
Notice that the usually reported membrane 3D viscosity is given by 
$\eta/h$, where $h$ is the membrane thickness.
A more detailed derivation of eq.~(\ref{hmeqn:6}) in terms of the 
total momentum decay is given in \cite{seki-93}.
Since 3D version of eq.~(\ref{hmeqn:6}) was originally used by 
Brinkman for flows in porous media~\cite{brinkman-47}, we call it 
as 2D Brinkman equation.       
But it should be kept in mind that there is no real connection 
between porous media and membranes except from the momentum decay 
mechanism.
For a typical flow in a membrane, we can adopt the Stokes approximation
to eq.~(\ref{hmeqn:6}).
Then we have 
\begin{equation}
\eta \nabla^2\mathbf{v} - \nabla p - \lambda \mathbf{v} = 0. 
\label{hmeqn:2}
\end{equation}
On the other hand, the incompressibility condition is given by
\begin{equation}
\nabla \cdot \mathbf{v}=0.
\label{hmeqn:8}
\end{equation}
Due to the presence of the momentum decay term, the Stokes paradox is 
now eliminated and eqs.~(\ref{hmeqn:2}) and (\ref{hmeqn:8}) can be solved 
analytically.

For a supported membrane, Evans and Sackmann assumed that a thin
lubricating layer of bulk fluid (thickness $H$ and 3D viscosity 
$\eta_{\rm f}$) exists between the membrane and
substrate~\cite{evans-88}.
In this case, they specifically identified the friction parameter as 
$\lambda=\eta_{\rm f}/H$ provided that $H$ is small enough.
Here we consider that the parameter $\lambda$ represents all kinds of 
frictional interactions between the lipid head group and the surrounding 
fluid even for a free membrane, and hence has more general significance.
Previously, the above Brinkman model was used to calculate velocity 
autocorrelation function of a disk~\cite{seki-93}, diffusion coefficient 
of a polymer chain in membranes~\cite{komura-95}, or dynamics of 
concentration fluctuations in binary fluid membranes~\cite{seki-07}.
As pointed out in ref.~\cite{oppenheimer-09}, the translational 
symmetry along the membrane surface is broken for eq.~(\ref{hmeqn:2})
due to the friction term.
We have implicitly assumed here that the velocity at infinity vanishes
so that the friction term is proportional to $\mathbf{v}$ itself.
We also note that eqs.~(\ref{hmeqn:2}) and (\ref{hmeqn:8}) describe 
a purely 2D model so that the 3D hydrodynamic nature of the bulk fluid 
is not taken into account.

\begin{figure}
\begin{center}
\resizebox{0.8\columnwidth}{!}{%
\includegraphics{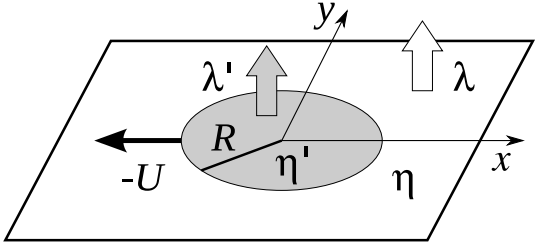}
}
\caption{Schematic picture showing a section of an infinite membrane 
with viscosity $\eta$ and momentum decay parameter $\lambda$ in which 
a circular liquid domain of radius $R$ with viscosity $\eta'$ and 
momentum decay parameter $\lambda'$ is moving.  
The liquid domain moves with a velocity $-U$ in the $x$-direction.
}
\label{schematic}
\end{center}
\end{figure}

As schematically presented in fig.~\ref{schematic}, we consider a 
circular liquid domain of radius $R$ and viscosity $\eta'$ moving with 
a velocity $-\mathbf{U}=(-U,0)$ in a thin membrane sheet of viscosity 
$\eta$.
As a result of the domain motion, a velocity field is induced around 
it as well as inside the domain.
Our purpose is to calculate the drag force experienced by the
domain. 
To generalize our treatment, we allow for the situation where the 
momentum decay parameters are different between the matrix  
($\lambda$) and the domain ($\lambda'$).
Hereafter, quantities without prime correspond to those of the matrix,
while quantities with prime refer to those of the domain.

The fluid velocity can generally be expressed as a gradient of some 
potential $\varphi$ and curl of some vector $\mathbf{A}= (0,0,A)$ which 
has only a single component~\cite{lamb-75}. 
Then the velocities in the matrix and the domain regions are 
expressed as 
\begin{equation}
\mathbf{v}= -\nabla \varphi + \nabla \times \mathbf{A}, 
\label{hmeqn:10}
\end{equation}
and
\begin{equation}
\mathbf{v}'=-\nabla \varphi' + \nabla \times \mathbf{A}', 
\label{hmeqn:10b}
\end{equation}
respectively.
Substituting eqs.~(\ref{hmeqn:10}) and (\ref{hmeqn:10b}) into 
eq.~(\ref{hmeqn:8}), we obtain
\begin{equation}
\nabla^2 \varphi = 0,~~~~~\nabla^2 \varphi' = 0, 
\label{hmeqn:11}
\end{equation}
which are the Laplace equations. 
With the use of eqs.~(\ref{hmeqn:10}) and (\ref{hmeqn:10b}), one can show 
that eq.~(\ref{hmeqn:2}) can be satisfied if the pressures are given by 
\begin{equation}
p=\lambda\varphi,~~~~~p'=\lambda'\varphi',
\end{equation}
while $A$ and $A'$ obey the following equations: 
\begin{equation}
(\nabla^2 - \kappa^2) A = 0,~~~~~
(\nabla^2 - \kappa'^2) A'= 0.
\label{hmeqn:13}
\end{equation}
Here we have defined the (inverse) hydrodynamic screening 
lengths for the matrix and the domain as 
$\kappa \equiv (\lambda/\eta)^{1/2}$ and
$\kappa' \equiv (\lambda'/\eta')^{1/2}$,
respectively.
We shall next seek for the solutions to eqs.~(\ref{hmeqn:11}) and 
(\ref{hmeqn:13}) subject to the appropriate boundary conditions for 
the translational motion of the domain.

\section{Calculation of drag coefficient}
\label{drag}

\subsection{Velocity and stress tensor}

It is convenient to work in the cylindrical polar coordinates 
$(r,\theta)$ with the origin at the center of the circular domain.
First we consider the matrix region where $r > R$.
Under the condition that the velocity and pressure must approach zero 
at large distances, we write down the solutions to eqs.~(\ref{hmeqn:11}) 
and (\ref{hmeqn:13}) as follows:
\begin{equation}
\varphi=\frac{C_1}{r}\cos\theta,~~~~~
A=C_2 K_1(\kappa r) \sin\theta.
\label{solutionout}
\end{equation}
Here, $C_1$ and $C_2$ are unknown coefficients which will be determined 
from the boundary conditions, and $K_1 (z)$ is the modified
Bessel function of the second kind of order one. 
Although the general solutions for $\varphi$ and $A$ can be expressed 
as series expansions in terms of $r$, we have kept only the least number 
of terms satisfying the requisite pressure and velocity conditions.
From eq.~(\ref{hmeqn:10}), the radial and tangential components of the 
velocity are given by 
\begin{align}
v_r&=\left[\frac{C_1}{r^2} + \frac{C_2}{r}K_1(\kappa r)
\right] \cos\theta, \\
v_\theta&=\left[
\frac{C_1}{r^2} + C_2 \kappa K_0(\kappa r) + 
\frac{C_2}{r}K_1(\kappa r)
\right] \sin\theta,
\end{align}
where we have used the recursion relations among the modified Bessel
functions (see Appendix A).
Then the components of the stress tensor can be obtained as
\begin{align}
\sigma_{rr}&= - p + 2\eta \frac{\partial v_r}{\partial r}
\nonumber\\
&=- \eta\left[ 
C_1 \left( 
\frac{\kappa^2}{r} 
+\frac{4}{r^3} \right) 
\right. \nonumber\\
&\left. 
+ C_2 \left( \frac{4}{r^2} K_1(\kappa r)
+ \frac{2\kappa}{r} K_0(\kappa r) \right) \right]
\cos\theta, \\
\sigma_{r\theta}&=\eta\left[ 
\frac{1}{r} \frac{\partial v_r}{\partial \theta}
+ \frac{\partial v_\theta}{\partial r}
- \frac{v_\theta}{r} \right]\nonumber\\
&= -\eta\left[ 
\frac{4C_1}{r^3}  \right.\nonumber\\
&\left.+ C_2 \left( 
\frac{4}{r^2}K_1(\kappa r)  
+\kappa^2   K_1 (\kappa r)
+\frac{2 \kappa}{r} K_0(\kappa r)
\right)\right] \sin\theta. 
\end{align}

Inside the liquid domain where $r<R$, the proper solutions to 
eqs.~(\ref{hmeqn:11}) and (\ref{hmeqn:13}) subject to the condition 
that they do not diverge as $r \rightarrow 0$ are
\begin{equation}
\varphi' =C_1'r\cos\theta,~~~~~
A'=C_2' I_1(\kappa' r) \sin\theta,
\label{solutionin}
\end{equation}
Here, $C_1'$ and $C_2'$ are unknown coefficients, and $I_1 (z)$ 
is the modified Bessel function of the first kind of order one. 
Since the corresponding radial and tangential components of the velocity 
are now
\begin{align}
v'_r&= \left[
-C_1'+\frac{C_2'}{r}I_1(\kappa' r)
\right] \cos\theta, \\
v'_\theta&=\left[
C_1' - C_2' \kappa' I_0(\kappa' r) + \frac{C_2'}{r}I_1(\kappa' r)
\right] \sin\theta,
\end{align}
the components of the stress tensor can be obtained as 
\begin{align}
\sigma'_{rr}&= - \eta' \left[
C_1'\kappa'^2 r
+ C_2' \left( \frac{4}{r^2} I_1(\kappa'r)
- \frac{ 2\kappa'}{r} I_0(\kappa'r) \right)  \right]
\cos\theta, \\
\sigma'_{r\theta}&= - \eta' C_2' \left[
\frac{4}{r^2}I_1(\kappa'r) + \kappa'^2  I_1(\kappa' r)
-\frac{2\kappa'}{r} I_0(\kappa' r) \right] 
\sin\theta. 
\end{align}

\subsection{Boundary conditions}

Next we assume that the no-slip condition is satisfied at the domain 
boundary.
This means that, at $r=R$, the radial component of the fluid velocities 
should 
be equal to the domain velocity $-U \cos\theta$, 
the tangential components should be continuous, and so should the 
components of the stress tensor~\cite{landaulifshitz-flumech}.  
These conditions are written as 
\begin{align}
\label{boundary1}
v_r &=-U \cos\theta, \\
v'_r &=-U\cos\theta, \\
v_\theta&=v'_\theta, \\
\label{boundary4}
\sigma_{r\theta}&=\sigma'_{r\theta}.
\end{align}
Since we have kept only the lowest order terms in $r$ for 
the solutions to eqs.~(\ref{hmeqn:11}) and (\ref{hmeqn:13})
(see eqs.~(\ref{solutionout}) and (\ref{solutionin})), we are allowed
to neglect the shape deformation of the circular domain while it moves.
In other words, we are implicitly assuming that the line tension 
at the domain boundary exceeds a typical viscous force.
A more quantitative argument to justify this condition will be given 
later.

\begin{figure}[t]
\begin{center}
\resizebox{0.85\columnwidth}{!}{%
\includegraphics{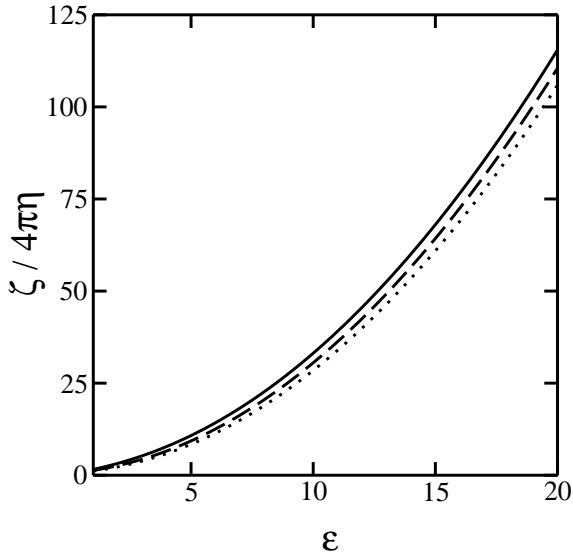}
}
\caption{The dimensionless drag coefficient $\zeta$ as a function of 
dimensionless domain radius $\epsilon=\kappa R$.
The different curves are for $E=0.1$ (solid), $E=1$ (dashed), and 
$E=10$ (dotted).}
\label{e10e1}
\end{center}
\end{figure}

Using the above four boundary conditions, we can obtain the 
four unknown coefficients $C_1, C_2, C_1'$ and $C_2'$ which are 
listed in Appendix B.
In the following, we introduce the dimensionless parameters to measure 
the relative viscosities $E \equiv \eta/\eta'$ and the relative 
decay parameters $L \equiv \lambda/\lambda'$.
We also use the notations $\epsilon \equiv \kappa R$ and 
$\epsilon' \equiv \kappa' R = \epsilon \sqrt{E/L}$ to measure 
the domain radius. 
Furthermore, the arguments of the modified Bessel functions 
will be omitted as $K_n \equiv K_n(\epsilon)$ and $I_n \equiv I_n(\epsilon')$  
in order to keep the notation compact.

\subsection{Force exerted on a domain}

The force exerted on the domain is given by the integral of the stress 
tensor over the boundary:
\begin{equation}
F= R \int_0^{2\pi} {\rm d}\theta \, 
(\sigma_{rr} \cos\theta - 
\sigma_{r\theta}\sin \theta).
\end{equation}
After some calculation, the drag coefficient obtained from $\zeta=F/U$
becomes 
\begin{align}
&\frac{\zeta(\epsilon; E,L)}{4\pi \eta} = 
\frac{\epsilon^2}{4} \nonumber\\
&+\frac{\epsilon K_1 [ (4+\epsilon'^2)I_1-2\epsilon'I_0+2E(\epsilon'I_0-2I_1)]}
{ K_0[(4+\epsilon'^2)I_1 - 2\epsilon'I_0] + E(2K_0+\epsilon K_1)
(\epsilon'I_0-2I_1)}.
\label{friction}
\end{align}
An alternate form can be obtained by using the following recurrence 
relation
\begin{equation}
I_0(\epsilon')= \frac{2 I_1(\epsilon')}{\epsilon'}+ I_2(\epsilon'),
\end{equation}
so that we have
\begin{equation}
\frac{\zeta(\epsilon; E,L)}{4\pi \eta} = 
\frac{\epsilon^2}{4} +
\frac{ \epsilon K_1 ( \epsilon' I_1 - 2 I_2+ 2 E I_2)}
{K_0(\epsilon' I_1 - 2 I_2) + E(2 K_0+\epsilon K_1)I_2}.
\label{frictionb}
\end{equation}
This is the main result of the paper.

\begin{figure}[t]
\begin{center}
\resizebox{0.85\columnwidth}{!}{%
\includegraphics{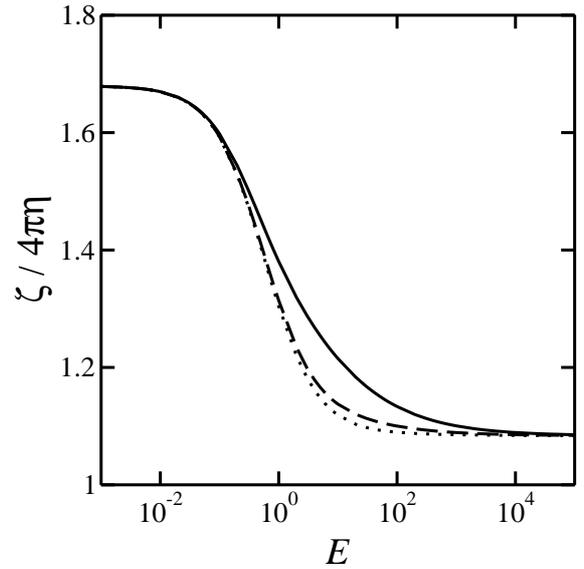}
}
\caption{The dimensionless drag coefficient $\zeta$ as a function of the
relative viscosity ratio $E=\eta/\eta'$ when $\epsilon=1$.
The different curves are for $L=0.1$ (solid), $L=1$ (dashed) and
$L=10$ (dotted).
}
\label{fnofe}
\end{center}
\end{figure}

\begin{figure}[t]
\begin{center}
\resizebox{0.85\columnwidth}{!}{%
\includegraphics{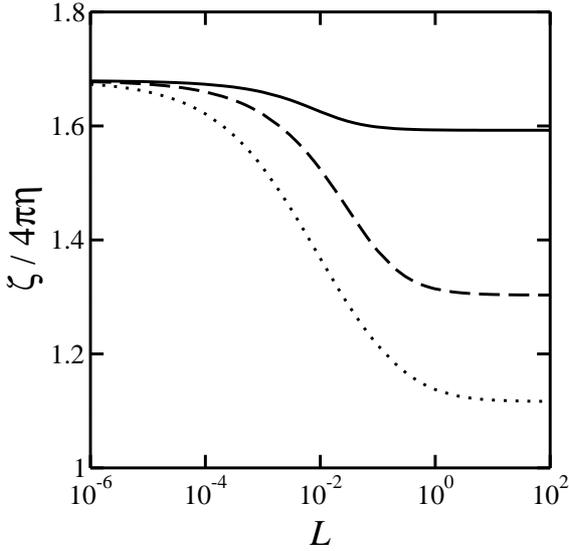}
}
\caption{The dimensionless drag coefficient $\zeta$ as a function of the
relative momentum decay parameter ratio $L=\lambda/\lambda'$ when 
$\epsilon=1$.  
The different curves are for $E=0.1$ (solid), $E=1$ (dashed) and
$E=10$ (dotted).}
\label{fnofl}
\end{center}
\end{figure}

In fig.~\ref{e10e1}, we plot the dimensionless drag coefficient 
$\zeta$ as a function of dimensionless domain size 
$\epsilon =\kappa R$ for $E=0.1, 1, 10$ when $L=1$. 
In all these cases, the drag coefficient increases with the domain 
radius $R$ as it should be.
For fixed values of $\epsilon$, on the other hand, the drag 
coefficient is smaller when the domain viscosity $\eta'$ becomes 
smaller (larger $E$). 
Fixing the domain size to $\epsilon=1$, we have plotted in 
figs.~\ref{fnofe} and \ref{fnofl} the drag coefficient $\zeta$ as 
a function of $E$ and $L$, respectively.  
In the former, we chose various values of $L$ ranging from 
$L=0.1$ to $10$, while the values of $E$ were changed from $E=0.1$ 
to $10$ in the latter.
Note that these graphs are both semi-log plots.
From fig.~\ref{fnofe}, we see that the drag coefficient 
monotonically decreases with increasing $E$ (smaller domain viscosity).
This can be attributed to the internal flows generated in the domain
because they are more efficient in transporting the fluid around it.
As a result, a domain with finite viscosity feels a smaller drag force
than a rigid disk.
The asymptotic values of $\zeta$ for $E \rightarrow 0$ and 
$E \rightarrow \infty$ converge to respective constants. 
In fig.~\ref{fnofl}, the drag coefficient is a decreasing function 
of $L$.  
The large $L$ values of $\zeta$ are dependent on $E$ as discussed below.

\subsection{Limiting expressions}

Next we examine several asymptotic limits of eq.~(\ref{friction})
or eq.~(\ref{frictionb}).
Some useful formulae are listed in Appendix A.

First we consider arbitrary values of $E$ and $L$.
In the limit of $\epsilon \rightarrow 0$, eq.~(\ref{friction}) gives a 
logarithmic behavior,
\begin{equation}
\frac{\zeta(\epsilon \rightarrow 0)}{4\pi\eta} \approx 
\frac{1 + E}
{\ln{(2/\epsilon)}-\gamma+  E [\ln{(2/\epsilon)}-\gamma+(1/2)]},
\label{xzero}
\end{equation}
where $\gamma = 0.5772 \cdots$ is Euler's constant. 
The above $E$-dependent logarithmic behavior is a new outcome
of our result.
In the opposite $\epsilon \rightarrow \infty$ limit, we have an 
algebraic dependence
\begin{equation}
\frac{\zeta(\epsilon \rightarrow \infty)}
{4\pi\eta} \approx \frac{\epsilon^2}{4},
\label{xinfinity}
\end{equation}
which does not depend on $E$. 
We also note that the above asymptotic expressions for small and 
large domain size limits do not depend on $L$.

\begin{figure}[t]
\begin{center}
\resizebox{0.85\columnwidth}{!}{%
\includegraphics{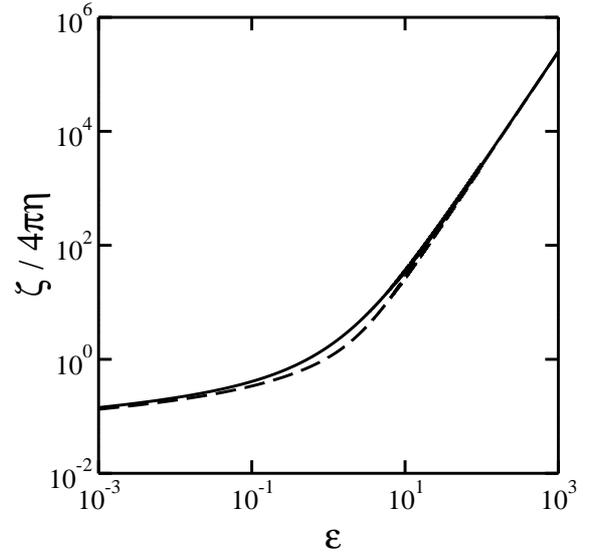}
}
\caption{The limiting dimensionless drag coefficient $\zeta_0$ (solid, 
eq.~(\ref{zetazero})) and $\zeta_{\infty}$ (dashed, eq.~(\ref{zetainfty})) 
as a function of dimensionless domain radius $\epsilon=\kappa R$ when 
$L=1$.  }
\label{ezeroinfty}
\end{center}
\end{figure}

For arbitrary $\epsilon$, we next consider the rigid disk case 
($\eta' \rightarrow \infty$) where the limit of $E \rightarrow 0$ 
is taken. 
Then we obtain
\begin{equation}
\frac{\zeta(\epsilon; E\rightarrow 0)}{4\pi\eta}= 
\frac{\epsilon^2}{4} + 
\frac{\epsilon K_1}{K_0}
\equiv \frac{\zeta_0(\epsilon)}{4\pi \eta},   
\label{zetazero}
\end{equation}
which coincides with the result by Evans and Sackmann \cite{evans-88}.
The opposite limit of $E\rightarrow\infty$ corresponds to a 2D 
gas bubble ($\eta' \rightarrow 0$). 
In this case, we obtain a different expression:
\begin{equation}
\frac{\zeta(\epsilon; E\rightarrow \infty)}{4\pi\eta}= 
\frac{\epsilon^2}{4} + 
\frac{2\epsilon K_1}
{2 K_0 + \epsilon K_1}
\equiv \frac{\zeta_{\infty}(\epsilon)}{4\pi \eta}.
\label{zetainfty}
\end{equation}
In fig.~\ref{ezeroinfty}, we plot both $\zeta_0$ and $\zeta_{\infty}$ 
as a function of $\epsilon$ in log-log scales. 
In the limit of $\epsilon \rightarrow \infty$, the difference between the 
two curves is the order of $\epsilon$ which becomes negligibly small
compared to the first terms in eqs.~(\ref{zetazero}) and
(\ref{zetainfty}).
Upon taking the $\epsilon \rightarrow 0$ limit of eqs.~(\ref{zetazero})
and (\ref{zetainfty}), we obtain 
\begin{equation}
\frac{\zeta_0(\epsilon \rightarrow 0)}{4\pi\eta} \approx 
\frac{1}{\ln{(2/\epsilon)}-\gamma},
\label{noslip}
\end{equation}
\begin{equation}
\frac{\zeta_{\infty}(\epsilon \rightarrow 0)}{4\pi\eta} \approx 
\frac{1}{\ln{(2/\epsilon)}-\gamma+(1/2)}.
\label{zerotangential}
\end{equation}
These are nothing but the $E\rightarrow 0$ and $E\rightarrow \infty$ 
limits of eq.~(\ref{xzero}).
We note that small $\epsilon$ behavior of $\zeta_0$ and 
$\zeta_{\infty}$ do not coincide in fig.~\ref{ezeroinfty}.
The limiting expressions of eqs.~(\ref{zetazero}) and (\ref{zetainfty})
for $\epsilon \rightarrow \infty$ coincide with eq.~(\ref{xinfinity}).

In the limit of $L \rightarrow 0$, which is the case of high momentum 
dissipation in the domain region, we recover eq.~(\ref{zetazero}) again;
\begin{equation}
\frac{\zeta(\epsilon;L\rightarrow 0)}{4\pi\eta}= 
\frac{\epsilon^2}{4} + \frac{\epsilon K_1}{K_0}.
\end{equation}
In the opposite $L\rightarrow \infty$ limit, we obtain 
\begin{equation}
\frac{\zeta(\epsilon;L\rightarrow \infty)}{4\pi\eta}=
\frac{\epsilon^2}{4}+
\frac{2 \epsilon(1+E)K_1}
{2(1+E)K_0 + \epsilon E K_1}.
\label{zetaLinfty}
\end{equation}
which depends on $E$ as observed in fig.~\ref{fnofl}. 
Equation~(\ref{zetaLinfty}) reduces to eqs.~(\ref{zetazero}) and 
(\ref{zetainfty}) for $E\rightarrow 0$ and $E\rightarrow \infty$,
respectively.

\section{Conclusion and discussion}
\label{conclusion}

In summary, we have obtained the drag coefficient of a circular liquid 
domain which has a finite viscosity using a 2D hydrodynamic 
equation with momentum decay or the 2D Brinkman equation.
We showed that the drag coefficient decreases as the domain viscosity 
becomes smaller with respect to the matrix viscosity.  
This is because the flow in the domain helps to transport the fluid 
in the surrounding matrix more efficiently. 
Our result shows an $E$-dependence in the small domain size limit
(eq.~(\ref{xzero})).
New limiting expressions for the $E\rightarrow\infty$ (eq.~(\ref{zetainfty})) 
and $L\rightarrow\infty$ (eq.~(\ref{zetaLinfty})) are also obtained.

Several points merit further discussion.
We first discuss the realistic value of $E=\eta/\eta'$ in the 
case of lipid domains.
Lipid domains are known to arise from a phase separation between 
a liquid-ordered ($L_{\rm o}$) phase rich in saturated lipid such 
as sphingomyelin and a liquid-disordered ($L_{\rm d}$) 
phase rich in unsaturated lipid~\cite{veatch-05}.  
Applying the pulsed field gradient NMR spectroscopy, 
Or\"{a}dd \textit{et al.} measured lateral diffusion coefficients 
of a single lipid molecule both in the $L_{\rm o}$ and $L_{\rm d}$ 
phases~\cite{oradd-05}. 
With the use of the SD logarithmic expression for the diffusion
coefficient, the 2D viscosities of these phases are estimated as 
$\eta_{\rm o}\approx 1.6 \times 10^{-9}$ Ns/m and 
$\eta_{\rm d}\approx 0.4 \times 10^{-9}$ Ns/m 
at 293 K 
(the membrane thickness is chosen to be 
$h \approx 3.8 \times 10^{-9}$ m). 
When the $L_{\rm o}$ phase forms a domain in the matrix of the 
$L_{\rm d}$ phase, the typical viscosity ratio would be 
$E \approx 0.2$.
In the opposite case where the $L_{\rm d}$ phase forms a domain,
the ratio tends to be $E \approx 4$. 
Hence, it is necessary to take into account the finite viscosity 
ratio between the domain and the matrix rather than regarding the
domain as a rigid disk.
Incidentally, sphingomyelin constituting the $L_{\rm o}$ phases  
has a large head group which can protrude out into the 3D fluid, 
whereas the $L_{\rm d}$ phase are devoid of such 
structures~\cite{simons-97}.  
It is therefore reasonable to assume that the momentum decay parameter
$\lambda$ experienced by the $L_{\rm o}$ and $L_{\rm d}$ domains are
different.

When imposing the boundary conditions as given from eq.~(\ref{boundary1})
to (\ref{boundary4}), we have assumed that the domain does not undergo 
any shape deformation.
This implies that the line tension should be large enough to overcome 
the viscous force of deformation.
In the case of lipid domains, the line tension was measured to be 
$\sigma \sim 10^{-12}$ N using domain boundary flicker 
spectroscopy~\cite{esposito-07}.    
For a flow of the order of $U \sim 10^{-6}$ m/s, a typical viscous 
force turns out be $\eta U \sim 10^{-15}$ N which is much smaller 
than $\sigma$. 
Hence it is reasonable to assume that the line tension is large 
enough to maintain the circular shape of the domains unless the 
temperature is very close to the critical point.
Different boundary conditions were used in order to consider the 
relaxation of deformed domains in a polymer monolayer~\cite{mann-95}.

The crossover from the logarithmic to algebraic behaviors of the 
drag coefficient takes place when $\epsilon \sim 1$ or 
$R \sim \kappa^{-1}$.   
A typical domain size which corresponds to this condition is roughly 
estimated below. 
As stated in sect.~\ref{hydrodynamic}, the momentum decay parameter 
for a supported membrane was originally identified as 
$\lambda=\eta_{\rm f}/H$, where $\eta_{\rm f}$ is the 3D viscosity 
in the outer fluid and $H$ is the thickness of a thin lubricating 
water which exists between the membrane and substrate~\cite{evans-88}.
In this situation, we have the hydrodynamic screening length  
$\kappa^{-1}=\sqrt{\eta H/\eta_{\rm f}}$.
Although we do not assume the presence of a solid substrate, the 
hydrodynamic screening length in the Brinkman approach would typically 
be $\kappa^{-1} \sim 10^{-7}$ m when $H=10^{-8}$ m, 
$\eta_{\rm f} \sim 10^{-3}$ Ns/m$^2$ and 
$\eta \sim 10^{-9}$ Ns/m. 
Domain size in the order of $R \sim 10^{-7}$ m is accessible in the 
experiments.

Within the present Brinkman model, the drag coefficient has a $R^2$ 
dependence in the large size limit as in eq.~(\ref{xinfinity}).
Such a behavior was indeed observed experimentally for Brownian 
motion of phase-separated domains on stacked-supported
membranes~\cite{kaizuka-04}.  
However, this dependence differs from the asymptotic analysis 
of the SD model according to which the drag coefficient increases 
linearly with $R$ (similar to a 3D sphere)~\cite{hughes-81}.
For ternary vesicles, Cicuta \textit{et al.} have measured the drag 
coefficient of a domain which is proportional to $R$~\cite{cicuta-07},
supporting the SD model. 
The large scale behavior of the Brinkman model and the SD model differs 
because the former is essentially a 2D model while the latter has a 
3D character. 
It should be also noted that the hydrodynamic screening length 
in the SD theory is given by $\nu^{-1} \equiv \eta/\eta_{\rm f}$ 
which is different from $\kappa^{-1}$.
In the present Brinkman approach, the hydrodynamic screening length 
is the geometric mean of $\nu^{-1}$ and $H$.

At this stage we mention that, based on the SD theory, an integral
form of the drag coefficient for a 2D liquid domain with $E=1$ was 
obtained by De Koker as~\cite{dekoker-02} 
\begin{equation}
\frac{\zeta_{\rm DK}}{4\pi\eta} = \frac{1}{4}
\left[ 
\int_0^\infty {\rm d}z \, \frac{J_1^2(z)}{z^2 (z+\delta)} 
\right]^{-1},
\end{equation}
where $\delta \equiv \nu R$ and $J_1(z)$ is the Bessel function of 
the first kind. 
We have estimated the above integral numerically. 
In the limit of $\delta \rightarrow 0$, the above expression gives 
a logarithmic dependence on $R$, in agreement with our result for 
small $\epsilon$ as seen in eq.~(\ref{xzero}).
In the opposite limit of $\delta \rightarrow \infty$, however, 
$\zeta_{\rm DK}$ is proportional to $R$ due to the 3D nature of 
the SD theory. 
As describe above, this is different from eq.~(\ref{xinfinity}) 
which gives the $R^2$ dependence.

Several authors considered a more general scenario in which a disk 
or a domain moves in a membrane being located at a finite distance 
$H$ from a fixed 
substrate~\cite{lubensky-96,stone-98,fischer-04,Tserkovnyak-06,inaura-08}.
Among these, Stone and Ajdari numerically solved the coupled dual integral 
equations which are obtained as solutions for the Stokes 
equations~\cite{stone-98}.
They showed that the Brinkman model gives a qualitatively good 
approximation over a wide range of membrane-substrate separations.    
One can explicitly show that the Brinkman and the SD models are 
recovered in the limit of $H \rightarrow 0$ and $H \rightarrow \infty$, 
respectively.

For comparison, we finally write down the analogous expression 
for the drag coefficient of a spherical drop having 3D viscosity 
$\eta'_3$ and radius $R$ moving in a fluid of 3D viscosity 
$\eta_3$~\cite{rybczynski-11,hadamard-11,landaulifshitz-flumech}
\begin{equation}
\frac{\zeta_3}{2 \pi \eta_3 R} 
= \frac{2E_3+3}{E_3+1},
\end{equation}
where $E_3=\eta_3/\eta'_3$.
In the solid sphere limit ($\eta'_3 \rightarrow \infty$), we obtain 
$\zeta_3 = 6 \pi \eta_3 R$ (the Stokes result).
In the opposite bubble limit ($\eta'_3 \rightarrow 0$), we have
$\zeta_3 = 4 \pi \eta_3 R$.
The reduction in the drag coefficient with the decrease in the drop  
viscosity $\eta'_3$ is similar to that seen for a liquid domain in 2D
as discussed before.
It is also important to point out that the effect of viscosity ratio
$E_3$ exists for all the drop sizes $R$ because there is no typical 
length scale for a 3D fluid.
This is in sharp contrast to our model in which the hydrodynamic 
screening length $\kappa^{-1}$ plays a crucial role.     
\\

We thank S. L. Keller and Y. Sakuma for useful discussions.
This work was supported by KAKENHI (Grant-in-Aid for Scientific
Research) on Priority Area ``Soft Matter Physics'' and Grant
No.\ 21540420 from the Ministry of Education, Culture, Sports, 
Science and Technology of Japan.

\renewcommand{\theequation}{A.\arabic{equation}}  
\setcounter{equation}{0}  
\section*{Appendix A. Useful formulae}
\label{appa}

According to ref.~\cite{abram-stegun}, the following recursion 
relations hold: 
\begin{align}
K_0'(z)&= -K_1(z), \\
K_1'(z)&= -K_0(z) -\frac{K_1(z)}{z}, \\
I_0'(z)&=I_1(z), \\
I_1'(z)&=I_0(z)-\frac{I_1(z)}{z}.
\end{align}
Here the primes indicate the derivative with respect to $z$.

In the limit of $z\rightarrow0$, we have 
\begin{align}
K_0(z)&\approx \ln\left( \frac{2}{z}\right) - \gamma +
 \frac{1}{4}\left[1+\ln\left( \frac{2}{z}\right) 
- \gamma\right]z^2 + \cdots, \\
K_1(z)&\approx \frac{1}{z}- 
\frac{1}{2}\left[\frac{1}{2}+\ln\left( \frac{2}{z}\right) 
- \gamma\right]z+ \cdots,\\
I_0(z)&\approx 1+\frac{z^2}{4} + \cdots,\\
I_1(z)&\approx \frac{z}{2}+\frac{z^3}{16}+ \cdots,\\
I_2(z)&\approx \frac{z^2}{8} + \frac{z^4}{96}+\cdots,
\end{align}
where $\gamma = 0.5772\ldots$ is the Euler constant.
In the limit of $z\rightarrow\infty$, we have 
\begin{align}
K_n(z) & \approx \sqrt{\frac{\pi}{2z}}e^{-z} 
\left[ 1 + \frac{4n^2-1}{8z} + \cdots \right], \\
I_n(z) & \approx \frac{e^z}{\sqrt{2\pi z}} 
\left[ 1 - \frac{4n^2-1}{8z} + \cdots \right].
\end{align}

\renewcommand{\theequation}{B.\arabic{equation}}  
\setcounter{equation}{0}  
\section*{Appendix B. Coefficients $C_1, C_2, C_1'$ and $C_2'$}
\label{appb}

The coefficients $C_1, C_2, C_1'$ and $C_2'$ are given by 
\begin{align}
C_1 &= -R^2U - 2 K_1 R U /(\kappa K_0)
\nonumber\\
&+2 K_1^2 R^2  \eta U(-2I_1 +I_0 R \kappa')/[ K_0(M_1+M_2)],
\\
C_2&= 4 I_0 R  (\eta-\eta')\kappa'U/[\kappa(M_1+M_2)]\nonumber\\
&- 2I_1 U [ 4 \eta - \eta'(4+R^2\kappa'^2) ]/[\kappa(M_1+M_2)],
\\
C_1'&= U+ 2 I_1 K_1 R \eta \kappa  U/ (M_1+M_2), \\
C_2'&= 2 K_1 R^2 \eta  \kappa U/(M_1+M_2),
\end{align}
with
\begin{align}
M_1&=
2I_0 K_0 R ( \eta - \eta')\kappa' + I_0 K_1 R^2 \eta \kappa \kappa',
\\
M_2&=
-2I_1 K_1 R \eta \kappa - 4 I_1 K_0 \eta  + I_1 K_0 \eta'(4+R^2\kappa'^2).
\end{align}
In the above, $K_n = K_n(\kappa R) = K_n(\epsilon)$ and 
$I_n = I_n(\kappa' R) = I_n(\epsilon')$ as denoted in the text.



\begin{thebibliography}{99}
\bibitem{dekoker}
Owing to the reviewer's comment, we recently came to know that, based 
on the SD theory, De Koker obtained an integral form of the drag force 
of a liquid domain which has the same viscosity as that of the matrix.  
Although this result was not published, we will compare it with our 
result in sect.~\ref{conclusion}.
\end{thebibliography}
\end{document}